\documentclass[11pt]{article}
\usepackage[margin=1.0in]{geometry}
\usepackage{graphicx}
\usepackage{epsfig}
\usepackage{amssymb,amsfonts,amsmath,amsthm}
\usepackage{mathrsfs}
\usepackage{epstopdf}
\usepackage{hyperref,color}
\usepackage{natbib}
\usepackage{setspace}

\pdfoutput=1

\begin{document}

\title{When the actual world is not even possible}
\author{Christian W\"uthrich\thanks{I am indebted to audiences at the GAP in Konstanz, the workshop `Bridging Metaphysics and Philosophy of Physics' in Rochester, {\em eidos} in Geneva, {\em The Metaphysics of Entanglement} project at Oxford, the Centre for Philosophy of Time in Milano, and the Urbino Summer School in Philosophy of Physics for discussions, as well as to Craig Callender, Jonathan Cohen, George Darby, Baptiste Le Bihan, Ghislain Guigon, Stephan Leuenberger, and Keizo Matsubara for comments on an earlier draft of this essay. Work on this project has been supported in part by the American Council of Learned Societies through a Collaborative Research Fellowship and in part by the John Templeton Foundation (grant `Space and Time After Quantum Gravity').}}
\date{12 June 2018}
\maketitle



\begin{abstract}\noindent
Approaches to quantum gravity often involve the disappearance of space and time at the fundamental level. The metaphysical consequences of this disappearance are profound, as is illustrated with David Lewis's analysis of modality. As Lewis's possible worlds are unified by the spatiotemporal relations among their parts, the non-fundamentality of spacetime---if borne out---suggests a serious problem for his analysis: his pluriverse, for all its ontological abundance, does not contain our world. Although the mere existence---as opposed to the fundamentality---of spacetime must be recovered from the fundamental structure in order to guarantee the empirical coherence of the non-spatiotemporal fundamental theory, it does not suffice to salvage Lewis's theory of modality from the charge of rendering our actual world impossible.
\end{abstract}

\section{Introduction}\label{sec:intro}

Various approaches to formulating a quantum theory of gravity either presuppose or entail that fundamentally, there is neither space nor time. Instead, space and time emerge from the more fundamental, non-spatiotemporal structure of quantum gravity very much in the way that tables and chairs emerge from the more fundamental, non-chairtabular structure of quantum particle physics. 

This may have cataclysmic consequences for metaphysics: some philosophical analyses of causation, laws of nature, persistence, personal and material identity, and even modality crucially seem to rely on the fundamental existence of space and time. For instance, David \cite{lew86} characterizes possible worlds as unified by the spatiotemporal relations among their parts but as spatiotemporally isolated from other possible worlds. If borne out, the disappearance of space and time would motivate a new---fatal---objection to Lewis's account of modality: his pluriverse, for all its ontological abundance, does not contain our world.

Apart from questioning the truth of theories denying the fundamental existence of spacetime, the Lewisian may respond in several ways to this shock. The more promising strategies consist either in questioning the empirical coherence of any theory denying the fundamental existence of space and time, or in arguing that Lewisian modality only requires the existence, but not the fundamentality, of space and time, or even in largely conceding the point and relaxing the conditions on the worldmate relation perhaps to just natural external relations. 

This essay contends that the first strategy fails, even though it unveils an important foundational task for the defender of a theory-sans-fundamental spacetime. In order to avoid the charge of empirical incoherence, and thus to salvage the possibility of its own epistemic justification, such a theory must be shown to admit emergent spacetime. Thus, in circumventing the first Lewisian response the assumption of the second response---that spacetime exists at some ontological level---is granted. However, this by no means entails that the second response succeeds. In fact, it is argued that the merely emergent existence of space and time comes at an unpalatably high cost to Lewis. Finally, the third strategy, though generally viable, will be shown to be so flexible as to drain almost any content from Lewis's condition of co-inhabiting the same possible world, and thus of being `worldmates', and to potentially sit uneasily with the metaphysics of the fundamental structure.

Although this essay focuses on the Lewisian conception of modality, its point is much more general:  theories in quantum gravity promise to have far reaching implications for metaphysics. These implications are difficult to circumvent on even just a mild form of naturalism. The minimal naturalism assumed throughout this essay asserts that no metaphysical thesis may be in manifest contradiction with facts established by our best science in the sense that the physically possible worlds are a subset of the metaphysically possible ones. On this assumption, then, no physically possible world can be metaphysically impossible; for otherwise, metaphysics would (a priori) deem impossible what physics affirms is possible. This essay considers metaphysical implications of this minimal form of naturalism if current research in quantum gravity is indicative of our future best science. 

\S\ref{sec:qg} explicates the need for a quantum theory of gravity and argues that such a theory will be our most fundamental theory of gravity, at least to date. In \S\ref{sec:causet}, I will show how spacetime may not be part of the fundamental furniture of the world by introducing the conceptually simple and clean case of causal set theory. \S\ref{sec:unify} acts as a reminder of the fact that it is spatiotemporal relations that unify and isolate worlds in Lewis's pluriverse, and \S\ref{sec:fail} parses out the trouble it encounters if those spatiotemporal relations are absent. \S\ref{sec:empinc}, \S\ref{sec:emerge}, and \S\ref{sec:natext} articulate and discuss the three most promising Lewisian strategies in responding to the challenge presented, respectively. I will argue that they all fail. I offer some brief conclusions in \S\ref{sec:conc}.

\section{Quantum gravity and fundamentality}\label{sec:qg}

Today, there are two incumbent theories in physics with a serious claim to be not just true, but {\em fundamental} theories: the standard model of particle physics and general relativity. The former describes the structure of what are (so far) the smallest scales at which physics makes reliable predictions and concerns the constitution of matter; the latter encodes the large-scale structure of our universe and its history. Both theories make eminently accurate predictions and both stand unrefuted, at least if evidence is restricted to direct tests of these theories. 

Yet not everything is well in fundamental physics. For starters, the standard model and general relativity stand in significant conceptual tension. The standard model radically reconceptualizes matter and energy from the way they figured in pre-quantum theories, but general relativity relies on these obsolete notions. General relativity proposes an equally radically novel understanding of space and time and their joint interaction with their energy and matter content, departing from the pre-relativistic physics of space and time presupposed by the standard model.\footnote{Or at least from the not {\em fully} relativistic assumptions made in the standard model.} This conceptual disunity is philosophically unattractive, but the clincher for their joint untenability is that there exist phenomena for which we have compelling reason to believe that their successful description must involve both quantum and relativistic, i.e., gravitational, effects. In other words, a theory is needed which commands the resources to {\em combine} the quantum with relativity. This is the theory the field of quantum gravity seeks to formulate, and I shall call a {\em quantum theory of gravity} any such theory with the explanatory ambition to account for these phenomena.

One might complain that the term `phenomena' is ill-chosen given that they concern, e.g., the physics of the very early universe and the hitherto unobserved evaporation of black holes. Of course, it remains true that no {\em observation}, which is unambiguously quantum-gravitational in the sense necessary to justify the need for quantum gravity, has knowingly been made to date. But while the case for the evaporation of black holes may be more tenuous, the reasons to believe that our universe started out in a very dense state and that this state can only be correctly captured by a theory attending to both quantum and gravitational effects are firmly anchored in our currently best physical theories, the standard model and general relativity. The argument which translates these reasons into a need for a quantum theory of gravity requires little more in terms of assumptions about the actual world we inhabit beyond these reigning theories. So for present purposes I shall assume that a quantum theory of gravity is needed.

A quantum theory of gravity will be a fundamental theory, at least more fundamental than any other theory in physics currently or previously held to be true. Presently, I use fundamentality to denote a relation between theories, partially ordering the true theories of physics, perhaps including merely approximately true theories, or perhaps even including theories which have, or had, some currency in science. The fact that a theory is more fundamental than another in no way entails that the first theory is fundamental {\em simpliciter}. The relation I am interested in here is thus more appropriately termed `relative fundamentality', although I will often suppress the qualification in what follows. Let us state what the relevant sort of fundamentality is. 

To {\em establish} whether a given pair of theories exemplifies the relation of relative fundamentality may be a highly non-trivial matter, particularly once one abandons the exclusive focus on physics in favour of a consideration of special sciences. For present purposes, I shall put aside this question and the more general debate on reductionism it invites; both deserve greater care than I can devote to them here. However, we can supply some abstract characterization of relative fundamentality. 

For the sake of the present argument, I rely on the assumption that the set of all hitherto relevant theories is partially ordered by relative fundamentality. Furthermore, I assume that relative fundamentality is irreflexive and hence that the partial ordering is strict. Since it is a partial ordering, it is of course transitive. Its irreflexivity and transitivity imply that it is asymmetric. 

Let us call the set of relevant theories $T$. If $T$ is finite, it contains a `minimal' element, i.e., a theory that is not less fundamental than any other theory in the set of theories considered.\footnote{If $T$ contains infinitely many theories, then an additional assumption that there is such a minimal element must be made. In that case, though, for $T$ to contain at least one minimal element it suffices that every totally ordered subset of $T$ has a lower bound in $T$. This follows, mutatis mutandis, from Zorn's lemma \citep{zor35}. For a subset $A$ of a partially ordered set $(B, \leq)$, an element $x$ of $B$ is a {\em lower bound} of $A$ just in case for all $a\in A$, $x\leq a$. To have a lower bound means that for any subset of theories which are totally ordered with respect to fundamentality, there exists a theory $t$ in $T$---and not necessarily in the subset---which is the most fundamental. Since the relevant ordering relation is irreflexive, $t$ is actually more fundamental, rather than just no less fundamental, than those in the subset considered.} To demand that $T$ contain at least one minimal element does not rule out that there exist several distinct theories with a justified claim to being the most fundamental. 

My insistence that there exists at least one minimal element of $T$ is usually given expression in the stipulation that the partial ordering of $T$ be `well-founded'. A binary relation which induces a partial order on its domain is {\em well-founded} just in case every non-empty subset of the domain has a minimal element with respect to the relation.\footnote{More precisely, a partial order is well-founded if and only if the corresponding {\em strict} order is induced by a well-founded relation. This precisification, however, adds nothing to the case at stake since `relative fundamentality', as stated above, is an irreflexive relation.} \footnote{Note also that in a subset consisting of just two theories, neither of which is more fundamental than the other, both of its elements, rather than none, are minimal.}  Well-foundedness results in the present case from the demand that ``all priority claims terminate.'' \citep[500]{sch03}  In other words, if from anywhere in the relevant set one starts asking the question whether there exists a theory which is relatively more fundamental than the one at hand, and then whether there exists a theory which is relatively more fundamental than {\em that} theory, etc, we must reach a negative answer within a finite number of steps.\footnote{The well-foundedness of a partial order may fail because there turns out to be an actually infinite tower of ever more fundamental theories. Ruling out this logical possibility comes at no great cost, since as finite beings, we could never entertain an actual infinity of theories.} 

My characterization of fundamentality differs from others prevalent in the literature.\footnote{And is arguably not shared by David Lewis, as Ghislain Guigon conveyed to me. This does not affect my argument against his account of modality below.} At least in the philosophical literature, fundamentality is usually understood in ontological or ideological terms rather than as a relation between theories. More specifically, fundamentality typically gets explicated by relations of ontological dependence obtaining among objects or structures or by mereological relations of parthood or by relations of supervenience holding among properties or some combination of these, e.g.\ in that properties of objects which ontologically depend upon, or are mereological complexes of, more basal or simple objects supervene on the properties exemplified by the basal or simple objects. The relevant relations are then taken to induce a partial ordering on their domain. To take theories to be the primary target of considerations regarding fundamentality instead of the objects they describe and their properties reflects my conviction that in fundamental physics theories indeed (epistemically) precede their ontological commitments and that it is thus more fruitful to address the question at the level of theories. Of course, the fundamentality relations as they obtain among theories will entail relations of ontological dependence among the objects or structures they postulate and relations of supervenience among the properties they ascribe to these objects or structures. In fact, one would hope that judgments about the fundamentality of theories are precisely mirrored by judgments concerning other ways in which fundamentality is considered. In what follows, we will be concerned with the fundamentality of an entity---spacetime. An entity will be considered fundamental just in case one among the most fundamental theories entails that it exists.\footnote{Clearly, this is debatable, e.g.\ with grounding realists. Since this is not the focus of my paper, I will refrain from doing so.}


Under the current description, the attempted quantum theory of gravity will certainly be more fundamental than general relativity if its ambition will be actualized. After all, this is its stated goal: to offer a theory of gravity which cannot only deal with the relativistic aspects of gravity, and hence of spacetime, but which also incorporates pertinent quantum effects. In other words, it endeavours to deliver a fundamental theory of gravity, and as such will be more fundamental than our currently most fundamental theory of gravity. Depending on the particular features of a candidate theory of quantum gravity, it may or may not be more fundamental than the standard model of particle physics. As stated above, the standard model offers our currently most fundamental theory of the three non-gravitational forces. Thus, if the quantum theory of gravity to be does not only provide a fundamental theory of gravity, but instead a unified theory of all forces, then it will also be more fundamental than the standard model. Such is the ambition harboured by string theory, for example. If, however, it only amounts to a fundamental theory of gravity, then it will not be more fundamental than the standard model. This will be the case, e.g., for most approaches trying to quantize general relativity such as loop quantum gravity and for most approaches starting out from more revisionary assumptions such as causal set theory. It should be noted, however, that it is also not the case that the standard model is more fundamental than these approaches. 

In either case, the quantum theory of gravity will be a minimal element of $T$ and thus among the most fundamental theories, at least to date. Let us proceed on this premise.

\section{The disappearance of space and time}\label{sec:causet}

According to most approaches to quantum gravity, spacetime is not part of the fundamental furniture of our world, but merely `emergent'.\footnote{\citet{hugwut13b,hugwut}.} What I mean by this is that whatever fundamental structure a theory of quantum gravity postulates, it is importantly dissimilar from the structure `spacetime' refers to in general relativity or other, non-fundamental theories of gravity or of spacetime. For instance, in a vast class of approaches to quantum gravity, the fundamental structure is discrete.\footnote{\citet[549]{smo09}.} In so-called canonical approaches to quantum gravity, there is a strong suggestion that at least time is not fundamental.\footnote{\citet[\S2]{hugeal13}.} Dualities in string theory may be interpreted to mean that space(time) is not fundamental in string theory either. The so-called Weyl symmetry of the `internal metric' already present in plain vanilla string theories is often interpreted, at least among physicists, as rendering unnecessary the background spacetime in which the string was at first assumed to propagate.\footnote{\cite{wit96}, but cf. also \cite[\S3]{hugeal13}.} Taking the theory ontologically seriously, as string theorists tend to, thus means accepting the vanishing of the spacetime at the fundamental level. In non-commutative geometry, an approach related to string theory, the fundamental constituents replacing the familiar spatiotemporal quantities in different dimensions do not commute, i.e.\ where measuring quantities along different dimensions depends on the {\em order} in which these measurements are made. Despite its spatiotemporal vestige, the non-commutativity of the approach renders the fundamental structure profoundly different from the ordinary conception of spacetime. And the examples could be multiplied. 

But rather than fully establishing the assertion that approaches to quantum gravity generically deny the fundamental existence of space and time, I shall content myself with offering a representative yet tractable example: causal set theory. Causal set theory is a still inchoate, but conceptually clean approach to quantum gravity, which may serve as a perfect illustration for how radically a discrete fundamental structure can differ from our usual conception of spacetime. Causal set theory is based on the assumption that the fundamental structure is a discrete set of featureless basal events partially ordered by causality. It is motivated by theorems in general relativity by Stephen \cite{haweal76} and David \cite{mal77} which establish that given the causal structure and some volume information, the metric of the spacetime manifold is determined, as is its dimension, topology, and differential structure. In other words, the causal structure determines the geometry, albeit not the `size' of the spacetime. Based on these theorems, causal set theory asserts that the fundamental structure is---or, more cautiously, is best represented by---a `causal set' and thus that causality is prior to space and time. Furthermore, the presupposition of the discreteness of the fundamental structure is justified through its technical and conceptual utility. 

Slightly more formally, causal set theory postulates that the fundamental structure is best represented by a {\em causal set} $\mathcal{C}$, i.e.\ an ordered pair $\langle C, \preceq\rangle$ consisting of a set $C$ of elementary events and a relation, denoted by the infix $\preceq$, defined on $C$ satisfying two conditions: first, $\preceq$ induces a partial order on $C$ (i.e., $\preceq$ is reflexive, antisymmetric, and transitive); second, $\mathcal{C}$ is discrete in that the number of elements of $C$ which are causally `between' any two points in $C$ is finite.\footnote{More technically still, the second axiom demands that $\forall a, b \in C,$ card$\{x\in C|a\preceq x\preceq b\} < \infty$.} That the discreteness is {\em stipulated} is not in itself a problem, as long as it is ultimately vindicated by the scientific success of the theory. Thus, it is a {\em feature} of the theory. 

Causal set theory is plagued by two major challenges. First, like other discrete relational approaches to quantum gravity, it suffers from what is known as the `entropy crisis', viz.\ that the vast majority of basic structures satisfying the above postulate cannot be approximated b 
y, or physically related to, a relativistic spacetime. In other words, for most causal sets in causal set theory, no spacetime even remotely resembling ours emerges from it. Second, causal set theory as it has been articulated so far is a classical theory---it does not take quantum interference effects into account. This is hardly satisfactory when the goal was to produce a {\em quantum} theory of gravity. The hitherto unfulfilled hope is to kill both birds with one stone: both problems would be solved if only one could formulate additional constraints on the causal sets which exactly pick those causal sets with general-relativistic pendants and simultaneously act as independently justified principles governing an appropriately quantum dynamics. Thus, the goal is to formulate the dynamics in some principled and physically motivated way such that the dynamics will `drive' the causal sets to those which do in fact approximate a general-relativistic spacetime.

It must be emphasized, however, that philosophers should not be misled by the presence of these difficulties to dismiss causal set theory altogether and much less to shrug off its basic tenet that the fundamental is discrete and thus dissimilar from spacetime as we find it in all prior and less fundamental theories. As pronounced above, the discreteness of the fundamental structure is quite common. In fact, it is so strongly expected that entire research programmes---and not just causal set theory---are based on the presupposition that the fundamental structure in quantum gravity---whatever else it might exactly turn out to be---is discrete. So like all other approaches to quantum gravity, causal set theory is not without its share of foundational problems. Pointing these out does not, however, make for an interesting objection against the argument about to be offered as the disappearance of spacetime at the fundamental level is quite generic in quantum theories of gravity. In this sense, the following argument does not depend on the truth of causal set theory in particular. Causal set theory permits an easily tractable illustration of much more general features of quantum theories of gravity. The argument only relies on the recognition that the non-fundamentality of spacetime is generically either an assumption or a consequence of quantum theories of gravity, at least for extant ones. 

Furthermore, it could be complained that all causal set theory would establish, if borne out, was that spacetime just looked a bit different from what we expected when we thought that general relativity was our best theory of it. Sure, the complaint admits, spacetime is not the continuum that general relativity taught us it was; instead, we learn that spacetime is a discrete structure. But why should such progress in learning about the properties of spacetime---if progress it is---have any consequences for metaphysics, the complaint rhetorically asks. 

If the point is simply to insist that we call `spacetime' whatever fundamental structure gives rise to space and time as experienced by ordinary humans, then we have no debate. But the complaint would be mistaken if it purported that the fundamental structure shares the essential properties of ordinary space and time and hence deserves the honorific title of `spacetime'. First, almost all physical theories taught today demonstrably rely on the assumption that space and time are (infinitely) extended continua, as do a number of metaphysical theories of (diachronic) personal, or generally material, identity, of causation, of laws of nature, as well as characterizations of determinism, etc. Thus, discrete spacetime would conceptually diverge from what is assumed to be evidently the case in all these theories and could no longer serve as their basis. Second, and more seriously, it takes hard---and controversial---work by mathematical physicists to even {\em define} the concepts necessary to attribute to discrete structures properties that we so routinely see exemplified by continuous spacetimes. The affine and differentiable structure evaporates, usual topological concepts become inapplicable, metric properties must be entirely revamped and redefined and dissertations are necessary to work out reasonable notions of the discrete correlate of the dimension of a smooth manifold. All these usual concepts developed to articulate the properties of continuous spacetimes, including, most importantly, metric properties of duration and length, become inapplicable. Worse, the discrete structures of causal set theory arguably do not possess metric properties at all.\footnote{Cf.\ the chapter on causal set theory in \citet{hugwut}.}

Just as it would be presumptuous to call phlogiston `oxygen' and attribute the differences simply to the ignorance of phlogiston theorists about what oxygen really is, it would be mistaken to label these discrete structures `spacetime', given these profound differences. Terms assume their meaning in the holistic context of the theories in which they operate; if these contexts shift radically, it makes no sense to insist on the same term. In fact, using homonyms for profoundly distinct concepts only invites misapprehensions. 

Finally, there are two concerns regarding the fundamentality of a quantum theory of gravity. First, one might worry that other fundamental theories might frustrate the conclusion that spacetime does not exist, fundamentally. As elaborated in \S\ref{sec:qg}, the requisite fundamentality is not jealous---there could be many minimal theories in the partially ordered set $T$ as the well-foundedness did not require uniqueness. So even if no other theory is more fundamental than our quantum theory of gravity, it may still be the case that this is also true of many other theories. In particular, as may be welcomed by some, this notion of fundamentality also permits the failure of reductionism in that the existence of entirely disparate subsets of theories, i.e., subsets of $T$ across which no fundamentality relations hold. For example, physical theories may be ordered according to fundamentality among themselves and likewise for biological theories, with relative fundamentality not exemplified by any physico-biological pair of theories. Closer to the case at hand, if our quantum theory of gravity is such as to remain silent about the constitution of matter, it can only hope to be more fundamental than general relativity, but not than the standard model. It is certainly conceivable that if there is a plurality of minimal elements in $T$, only one of which is our quantum theory of gravity, then different minimal elements may return different verdicts regarding the status of spacetime. So even if our quantum theory of gravity declared that there is no spacetime, other theories might postulate its existence and by virtue of their fundamentality insist that it does so fundamentally. How to regulate the jurisdictions in such a case?

In response to this worry, the first point to note is that whatever set of minimal theories we have, it must be consistent. I take this not to imply that different elements of $T$ cannot assert different, and perhaps even inconsistent, propositions. What it does imply at a minimum, however, is that should such a case arise, then not both theories can be strictly true. At best, one of them is true, and the other one only approximately so. This is a matter of logic, and hence no substantive constraint on $T$. It means that one theory takes precedence over the other, at least with respect to the claims at stake. It also means that we should ideally possess a procedure to adjudicate the dispute over conflicting jurisdictions. Unfortunately, I know of no principled way of doing so generally. But we do not need a generally valid procedure; it suffices to provide an argument that a quantum theory of gravity's pronouncements regarding the fate of spacetime trumps those of other fundamental theories. And such an argument is readily available: our hitherto most accurate theory of spacetime is general relativity, our quantum theory of gravity is more fundamental than general relativity, so the nature of that fundamental structure which gives rise to relativistic spacetimes is most authoritatively described by the quantum theory of gravity. 

A second worry someone might voice trades on the fact that even though fundamentality is intended without regard of the contingencies of the current state of science, it may be that as yet unformulated theories will revert the verdict on spacetime as it is handed to us by extant quantum theories of gravity. Regardless of how many more fundamental levels will be uncovered, the possibility which instigates the present worry forces me to stipulate that at the most fundamental one or, in the case of infinitely many, all of which are more fundamental than {\em some} level do not reverse the outcome of \S\ref{sec:causet}, viz.\ that fundamentally, there is no spacetime. As long as this condition is honoured, the contingencies of future science do not affect the argument below. I believe that current research in fundamental physics warrants the tentative imposition of this condition and thus underwrites that there is no spacetime at the fundamental level, but I also understand that this warrant is, like all science, fallible. 

Let us then, at least for the purposes of this essay, accept that spacetime does not exist, fundamentally.

\section{What unifies a Lewisian world}\label{sec:unify}

Assuming familiarity at least with the basic ideas of Lewis's account of modality in terms of possible worlds as it is articulated in his \citeyear{lew86} et passim, let me remind you of what Lewis claims unifies possible worlds. In the section entitled `Isolation' (1986, \S 1.6), he addresses the question of by virtue of what two possibilia are `worldmates', i.e.\ denizens of the same possible world, and thus of what unifies and isolates possible worlds. For him, possible worlds are unified by the spatiotemporal relations holding among possibilia, and are distinct by virtue of being spatiotemporally isolated from one another. In his own words,
\begin{quote}
... things are worldmates iff they are spatiotemporally related. A world is unified, then, by the spatiotemporal interrelation of its parts. There are no spatiotemporal relations across the boundary between one world and another; but no matter how we draw the boundary within a world, there will be spatiotemporal relations across it. \citeyearpar[71]{lew86}
\end{quote}
Lewis himself identifies three problems with his account.\footnote{In fact, Lewis briefly addresses a fourth worry concerning the possibility of spirits living outside of space altogether. I shall ignore it here.} These problems, or perhaps rather {\em limitations}, of the account concern the necessity of the condition much more so than its sufficiency. First, it does not permit a world to ``consist of two or more completely disconnected spacetimes.'' (71) This limits Lewis's account insofar as a world which can be represented as the union of at least two disjoint non-empty spacetimes is deemed impossible. The relevant sense in which these spacetimes are disjoint, of course, must be that no spatiotemporal relations are exemplified by relata located in topologically separate parts of this world. As Lewis admits, and regrets, the impossibility of such worlds, he offers substitutes, i.e.\ possible worlds which exhibit some features emulating what a proponent of the possibility of disconnected spacetimes might have cared about. For instance, worlds might possess additional, covert, spatiotemporal dimensions along which seemingly disconnected regions of spacetime connect; or otherwise disjoint regions might be connected by virtue of their being embedded in a single spacetime such that topological or metric oddities effectively cordon off the regions from one another. 

This limitation might worry a naturalist who takes recently popular multiverse proposals metaphysically serious. However varied these proposals may be, they all essentially defend the theses that our universe is just one among many, and that these alternate universes are causally disconnected from ours. If this were true of our world---and if there indeed are no spatiotemporal relations obtaining across universes in a way that cannot be cured by substitutes---, then Lewis's account of modality would rule out our actual world as impossible. The trouble with this conditional, unlike the one defended in the main argument of this essay, is that I find little reason to accept either---let alone both---of the antecedents. While an analysis of the evidential claims made in favour of multiverse hypotheses would require an entirely separate paper, the more plausible versions of it do not sanction the second antecedent as they assume that all the universes, as causally separate as they may be, are bubbles embedded in a larger spacetime, connecting them in the presently relevant way. I conclude that this limitation does not pose a serious threat to Lewis's analysis from a naturalistic perspective.

The second issue with his analysis, on Lewis's own view, is that it is impossible that there is nothing. There needs to be at least a tiny bit of ``homogeneous unoccupied spacetime, or maybe only a single point of it'' (73) for there to be a world at all.\footnote{In fact, an empty world would be compatible with the present analysis of the worldmate relation; the incompatibility arises from Lewis's analysis of what a world is. I thank Stephan Leuenberger for this point. As Baptiste Le Bihan rightly said in correspondence, it is not possible that there is nothing on Lewis's view.} Lewis accepts that on his view, it comes out as a necessary truth that there is something rather than absolutely nothing at all. However, he repudiates that this {\em explains} in any way why this is so, as for him, explanation is etiological, i.e., in the business of providing a causal account, as is evidently not the case here. So long as this consequence is not mistaken as an {\em explanation}, Lewis is happy to accept it as a feature of his analysis, and so am I.

Thirdly, and most disconcertingly for Lewis himself, it seems as if a world in which events are related by two distinct types of `distance'---temporal and spatial---such as a Newtonian world ought to be possible, even though events in it are not related by ontologically fundamentally distinct {\em spatiotemporal} relations, as demanded by the condition. Of course, one might also want to include worlds in which, e.g., spatial distances are not naturally part of the story of how this world is interrelated, but are instead replaced by three distances: that along the left-right axis, along the up-down axis, and along the front-back axis. It might just turn out that these distances are essentially different in this world and in others like it, thus resulting in four, and not just two, distances by which objects in those worlds are related. Since there is no good reason to think that such worlds are impossible, the generalized version of this complaint continues, to demand that worlds are spatiotemporally bound is overly narrow. What we dub `spatiotemporal relations' is grounded in their behaviour in the actual world. The question which then arises is whether those other multiple distance relations obtaining in these other worlds are really nothing but our spatiotemporal relations which ``double up'' to deliver several distances, or whether the latter are different relations altogether, which ``take the place'' of our spatiotemporal relations. If the former, we need not worry, Lewis assures us---quite correctly in my view. If the latter, however, a can of worms in the metaphysics of relations is opened. He offers to deal with this objection by accepting that 
\begin{quote}
[w]hat I need to say is that each world is interrelated (and is maximal to such interrelation) by a system of relations which, if they are not the spatiotemporal relations rightly so called, are at any rate analogous to them. (75)
\end{quote}
He then goes on to state some preliminary points of analogy these `analogically spatiotemporal' relations must satisfy. The relations, at a minimum, must be `natural' (i.e.\ not gerrymandered or disjunctive), `pervasive' (i.e.\ if relata are connected by a chain of relations, they are also directly so related), `discriminating' (i.e.\ in sufficiently large worlds, the relata are possibly identified by a unique place in the structure of relations), and `external' (i.e.\ not supervenient on the intrinsic natures of the relata). Lewis also considers, but ultimately rejects, a simplification circumventing the difficulty of having to deal with this messy business by relaxing the condition that worlds must be unified by at least analogically spatiotemporal relations to the condition that they can be unified merely by virtue of {\em some} natural external relation. However the further details of the metaphysics of analogically spatiotemporal relations in Lewis's pluriverse look like, let me emphasize the dilemma Lewis faces at this juncture. Either the candidate relations (such as Newtonian spatial and temporal distances) unifying what he takes to be a nomologically distant world are nothing but our spatiotemporal relations behaving somewhat differently; or else they really are non-spatiotemporal, in which case the world they are supposed to unify is only possible if they are {\em analogically} spatiotemporal. We will return to the issue of relaxing the worldmate relation to be based on something weaker than spatiotemporal relations in \S\ref{sec:natext}.

\section{Unification failure}\label{sec:fail}

But now another problem arises in the light of the findings in \S\ref{sec:causet}. If borne out, the disappearance of space and time would motivate a new and, I would claim, fatal objection to Lewis's account of modality: despite its modal {\em embarras de richesses}, his pluriverse does not contain our world. In other words, if all there is is contained in the pluriverse of all possible worlds, {\em then the actual world would not exist}, as it would not be a possible world. This follows immediately from the fact that in no possible causal set, we have at least a tiny bit of ``homogeneous unoccupied spacetime''. The basic constituents of a causal set stand in {\em causal}, but not {\em spatiotemporal}, relations to one another. This seems to leave open the possibility of a world consisting of just one basal event, which perhaps could ``double up'' as a single point of spacetime. By virtue of what would this single event be a single point of spacetime? Spacetime, according to causal set theory, is an emergent, not a fundamental entity. But even if a single basal event doubling up as a point of spacetime is granted for the sake of argument, it seems as if at most one basal and structureless event populates a possible world as no spatiotemporal relations would be admitted fundamentally which could bind them together as worldmates. 

In other words, all the basic elements of a causal set must live in different possible worlds according to the Lewisian condition, or so it seems. If, per the usual assumption in causal set theory, one such element of the fundamental structure corresponds to a Planck-sized volume of spacetime, then the observable part of our actual universe corresponds to something like a gargantuan $10^{185}$ basic elements. If there is no doubling up or analogically spatiotemporal about such a causal set, then it appears that these elements must constitute numerically distinct worlds. In order to form something like the causal set representing the fundamental structure of our actual universe, an even more gargantuan number of causal relations must obtain across worlds. But Lewis rejects trans-world causation, and quite plausibly so. This injunction of causal isolation leaves us with a large number of possible worlds populated by just one basal event. But since these basal events obtain their identity only structurally,\footnote{Cf.~\citet{wut12c}.} we stand bereft of any power to ground their identity and hence their numerical plurality: since they are intrinsically indiscernible, they would all represent one and the same world.

Whether all these $10^{185}$ elements give rise to numerically distinct possible worlds, or whether they are numerically one and the same shall not concern us here. What {\em is} pertinent is the fact that a possible world containing nothing but a single basal element of a causal set looks nothing at all like our actual world. Hence, the actual world, if fundamentally as causal set theory asserts, is not a possible world in the Lewisian pluriverse.

Or at least not if the causal relations exemplified in the fundamental structure neither double up as spatiotemporal relations nor are analogically spatiotemporal. We can safely dismiss the first possibility: the causal relations featured in causal set theory are expressly {\em not} the spatiotemporal relations just behaving a little differently. They {\em give rise} to spatiotemporal relations, as we will see below, but differ markedly in both their extensional properties as well as their intensional role in the theory, as explained in \S\ref{sec:causet}. Thus, we are left with the option that the causal relations at work in the fundamental structure are analogically spatiotemporal, as will be more fully discussed in \S\ref{sec:extnat} below.\footnote{It should also be added that even though Lewis also considers causal relations, and particularly the possibility of their obtaining across worlds, he does not entertain them as solely sufficient to bind worlds. Lewis also declines to introduce a primitive worldmate relation vested with all the necessary unifying power.}

If the causal relations present in the fundamental structure of our world as conceived by causal set theory neither double up with spatiotemporal relations nor are they analogically spatiotemporal, then Lewis's account of modality, and perhaps more, is under threat. Since it is necessary, and not merely sufficient, for worldmates to stand either in spatiotemporal relations or in external relations which double for spatiotemporal relations or in analogically spatiotemporal relations, non-trivial causal sets would not constitute possible worlds in Lewis's sense. Therefore, because what is not possible cannot exist---for both the actualist and the modal realist---we may have a new problem: the non-existence of the actual world.\footnote{The reader is reminded that we proceeded in \S\ref{sec:causet} on the assumption that causal set theory or a relevantly similar theory is a true fundamental theory of our world.} Instead of many non-actual possible worlds, we have a non-possible actual world! 

How can the Lewisian respond to this challenge? Let us distinguish five responses:
\begin{enumerate}
\item deny the truth of causal set theory,
\item argue that even though fundamental structure may look quite dissimilar from spacetime as we know and love it, this structure is what spacetime looks like, fundamentally, and so there is fundamental spacetime;
\item question the empirical coherence of any theory denying the fundamental existence of spacetime;
\item argue that Lewisian modality only requires the mere existence, but not the fundamentality, of spacetime; and 
\item weaken the requirements on relations which ground worldmate relations, e.g.\ to natural external relations.
\end{enumerate}
The first two strategies for responding are unpromising. First, the implication that spacetime does not exist fundamentally is not specific to causal set theory; instead, it is quite generic in quantum gravity. In order to successfully complete this response, the Lewisian would have to deny the truth of any such theory of quantum gravity at any possible world. As for the second strategy, if, against what was argued above, `spacetime' just becomes a designation for whatever fundamental structure there is, then Lewis's principle of what makes worldmates becomes empty. 

The remaining three responses are much more promising and will be discussed, in turn, in the remaining sections. I will argue that however promising, they nevertheless ultimately fail.

\section{The charge of empirical incoherence}\label{sec:empinc}

Let me explain, then, how the absence of spacetime may incur a worry about the empirical coherence of a theory-sans-spacetime. If one believes, with Tim \cite{mau07}, that any realist interpretation of quantum mechanics must include `local beables' in its ontology, then one might be tempted to charge any theory denying the fundamental existence of spacetime with empirical incoherence. Coined by John \cite{bel87}, `local beables' in quantum mechanics are basic existents	``which are definitely associated with particular space-time regions.'' (234) They represent the physical content of the universe, they make up the `stuff' we find in our world. Whatever else their properties, \cite{mau07} makes it explicit that they require space and time when he asserts that ``local beables do not merely exist: they exist somewhere.'' (3157) Now abstracting away from non-relativistic quantum mechanics, insisting that the basic ontology of a theory includes local beables thus entails a commitment to spacetime itself being a part of the basic furniture of the world (and hence not emergent at a higher level). Whatever else confirmation in an empirical theory involves, the observation of something located in some place at some time will play an ineliminable role in it. 

Thus, it appears as if denying such fundamental existence to spacetime threatens a theory's `empirical coherence' in the sense of Jeff Barrett, who defines a theory to be {\em empirically incoherent} just in case the truth of the theory undermines our empirical justification for accepting it as true \citep[\S4.5.2]{bar99}. A theory denying the fundamental existence of spacetime is thus alleged to be empirically incoherent because the empirical justification of a theory derives only from our observation of the effects of local beables situated in spacetime; yet if such a theory were true, there could be no such local beables. 

To be sure, \citet[3161]{mau07} accepts---at least in the context of the nonrelativistic quantum mechanics he is concerned with---that it would be sufficient in principle to {\em derive} the structure of the local beables and the spacetime that contains them from a fundamental ontology free of spacetime and hence of local beables. Such a derivation would permit an empirical grounding of the theory-sans-spacetime, {\em as long as it is ``physically salient (rather than merely mathematically definable).''} (ibid.; emphasis added) Maudlin implies, however, that we do not even {\em know how} such a ``physically salient'' derivation could be undertaken, let alone having it at our disposal.\footnote{For a general strategy circumventing Maudlin's worry in cases of theories without fundamental spacetime, cf.~\citet{hugwut13b} and \citet{lamwut18}.}

To summarize then, a necessary condition to circumvent the charge of empirical incoherence is to offer a ``physically salient'' derivation of spacetime from the basic, non-spatiotemporal structure. In other words, it must be shown how general-relativistic spacetime emerges as an approximation in quantum theories of gravity. The task of recovering the smooth relativistic spacetime from the discrete fundamental structure is also important in the `context of justification', as its discharge provides an account of why the classical spacetime theory---general relativity---is as successful as it is. Understanding how spacetime emerges is thus doubly urgent: to establish the fundamental theory's empirical coherence by connecting it to evidence in the form of spatiotemporally located beables, and to provide an explanation of the successes---and failures---of the predecessor theory by relating the fundamental structures to relativistic spacetimes. 

The necessity of recovering spacetime in a well-understood approximation arises for every quantum theory of gravity which denies its fundamental existence. As I have asserted above, this applies to most approaches in quantum gravity. How exactly spacetime can be recovered depends on the particular approach taken and remains an open question for many of them. I have explicated the general idea for completing this task---cum attendant difficulties---as it appears in causal set theory in \citet{wut12c} and in \citet{hugwut}. For the present purpose, however, let us assume that the endeavour is both acknowledged and can be successfully completed. Although this outcome thwarts the third Lewisian response, it thereby grants the presumption of the fourth, to which we now move.

\section{Emergent salvation?}\label{sec:emerge}

Given the outcome of \S\ref{sec:empinc}, the existence of spacetime is assured, albeit not at the fundamental level. Spacetime, just like tables and chairs, is non-fundamental, but existent. So we can assume, at least for the sake of argument, that the non-fundamentality does not entail the non-existence of spacetime, just as it didn't for tables and chairs.\footnote{One can of course object to the existence of non-fundamental entities, e.g.\ (but not only) by denying the existence of mereological sums. Clearly, the present strategy will not be appealing to a Lewisian of this stripe. Thanks to Baptiste Le Bihan for pushing me on this point.}  In fact, by acknowledging the necessity of a ``physically salient'' emergence of spacetime, its existence was accepted, at least in an appropriate approximation. The needed spatiotemporal relations exemplified by pairs of worldmates, the thought goes, may well be ontologically dependent upon some more fundamental structure. Consequently, their relata need not, and in general will not, be fundamental objects either. 

Of course, this existential acknowledgement would only save the Lewisian if the non-fundamental existence of spacetime were sufficient to unify (and isolate) worlds. Unfortunately, this is not the case. There are three principal reasons why this emergent salvation of Lewis's theory of modality fails, each sufficient to bury any hope of salvation. 

First, a metaphysician will hardly be satisfied by being offered what physicists call an `effective theory' of the world, i.e.\ an at best approximately true theory which ignores the fundamental reality of that world in favour of a description of merely emergent phenomena. One might have thought that the argument in \S\ref{sec:empinc} established that any fundamental theory must be compatible with a theory of emergent spacetime {\em which is true simpliciter} in order to guarantee its empirical coherence. After all, how could a false theory ground our empirical judgments? Unfortunately, this will simply not be borne out: the relevant theory of emergent spacetime is general relativity, and general relativity's failure in regimes with strong gravitational fields where quantum effects matter was the very reason a quantum theory of gravity was sought in the first place. Thus, general relativity will be false in its pronouncements in these regimes. And it will only be approximately true in more placid domains. A remarkably accurate approximation, but an approximation nevertheless. Just as most of our statements about tables and chairs are true simpliciter, many statements concerning the observable domain entailed by general relativity will be true simpliciter. But just as in our folk theories of tables and chairs, not all will, frustrating the theory's truth simpliciter. So that's the non-fundamentalist Lewisian's first problem: worlds are unified by relations furnished by theories which are strictly speaking false. 

Second, although the actual world would come out as metaphysically possible, all those many causal sets which do not approximate a relativistic spacetime come out as impossible---even though a bona fide theory of physics asserts that they are (physically) possible. On the one hand, this would not be much of a loss, since none of these causal sets can describe the fundamental structure of our actual world, on pain of empirical incoherence. On the other hand, this would have the odd consequence that the theory correctly describing the fundamental structure of our world asserts nomological possibilities which are not metaphysically possible. Usually, the metaphysically possible worlds are considered a (proper) superset of the physically possible worlds. This would no longer be the case on this proposal. More problematically, the existence of metaphysically impossible physical possibilities would clearly violate the mild form of naturalism assumed at the outset. 

Third, a Lewisian theory of modality trading in emergent spatiotemporal relations must remain impotent in binding the basic constituents of actuality into one and the same possible world. The reason for this impotence is simple: there are no spatiotemporal relations exemplified in the fundamental structure of a causal set, and the basal elements of this structure cannot be the relata of the emergent spacetime relations. 

This latter claim needs explicating. If spatiotemporal relations did obtain between elementary events, then propositions asserting corresponding facts such as `This event and that event are at such-and-such spatiotemporal distance' would be meaningful. But metrical (and many topological etc.) concepts are inapplicable at the fundamental level of causal sets, as was argued in \S\ref{sec:causet}. It is notoriously difficult to define geometric notions in purely fundamental terms, i.e., in terms of causal set theory. There are various difficulties. First, even if successful, we should  expect that our familiar geometric notions will only approximately track the fundamental magnitudes, even if the analogy is as strong as it can be. Secondly, even for those magnitudes for which the fit is reasonably close, we should expect that they will give out in exactly those regimes for which general relativity was found wanting. Thirdly, many of the fundamental notions defined in the literature rely in their ability to successfully emulate emergent geometric notions on additional substantive assumptions about either the fundamental causal set or the emergent spacetime, or both. For instance, the notion of spatial distance has been plagued by all these difficulties \citep{ridwal09}. The only somewhat promising ansatz to define a notion which gives rise to something like a spatial distance in the literature (offered in \citealt{ridwal09}), can only (approximately) reproduce distance in Minkowski spacetime, but not in curved spacetimes. Since general relativity insists that our actual world is best understood as having a curved spacetime structure, this proposal does not help in saving the actual world from the impending impossibility.

One might be tempted to impose, by a strong hand, spatiotemporal relations to obtain between the basal events of a fundamental causal set, as follows. Suppose we have a relativistic spacetime and the fundamental causal set it approximates. Showing that the spacetime emerges from the causal set essentially amounts to finding a map $f$, satisfying certain conditions of well-behavedness that need not concern us here, from the causal set to the spacetime.\footnote{See the chapters on causal set theory in \citet{hugwut}.} For any two basal events $a$ and $b$ in the causal set, $f(a)$ and $f(b)$---being elements of the spacetime---surely stand in some spatiotemporal relation to one another. Since $f$ is injective, i.e., every element of its domain is unambiguously mapped onto an element of its range, it seems as if this move is available for any pair of basal events of any causal set which has any prayer of giving rise to an emergent spacetime (thus bracketing the second reason above). Since by definition no spatiotemporal relations are defined at the fundamental level of a causal set, couldn't one introduce surrogate spatiotemporal relations by stipulating that any two events $a$ and $b$ of a causal set stand in the spatiotemporal relation $R$ just in case the two spacetime points $f(a)$ and $f(b)$ stand in $R$? 

Thus, one would start out from a fundamental causal set capturing the fundamental structure of a world such as ours, find the relativistic spacetime which emerges from it, and then extend the spatiotemporal relations as they are defined of the spacetime points to obtain of pairs of basal events, as suggested in the previous paragraph. Unfortunately, as straightforward as this procedure appears, it is marred with insurmountable difficulties. First, it turns out that most causal sets admitted by the basic axioms of the theory cannot be embedded into a spacetime, as stated in \S\ref{sec:causet}. For all these causal sets it would still be the case that they do not constitute possible worlds on the Lewisian analysis of modality and the second reason above would kick in.

A second difficulty of the proposal to extend the spatiotemporal relations to causal sets from which a spacetime emerges arises from our assumption of the unique existence of the embedding spacetime. Uniqueness may be granted, but the problem is, once again, that relativistic spacetimes only approximate causal sets. In exactly those domains for which a quantum theory of gravity proved necessary, it will not be possible to map the causal set into a spacetime which is also a model of general relativity. This should be expected to happen exactly in those circumstances general relativity gives out. It is simply not nomologically possible, as far as general relativity is concerned, to have a spacetime which faithfully describes our world in these regimes. And note that these domains are believed to be part of our actual world. So we should not expect to find a relativistic spacetime which emerges globally from the causal set and truthfully describes the fundamental structure of our world. 

Thus, if causal set theory is true of our world, then the basic elements of reality do not stand in spatiotemporal relations and thus cannot populate the same possible world in Lewis's analysis.  Given our failed attempts to precisely identify {\em spatiotemporal} relations at the fundamental level, couldn't the causal relation, which {\em is} fundamentally exemplified, be reinterpreted as a relation of {\em temporal precedence}? After all, it partially orders the elementary events, which is exactly what time does in special relativity. But this escape falls short on two counts. First, the motivating theorems by Malament and others mentioned in \S\ref{sec:causet} strongly suggest an interpretation of the fundamental relation as being closely related to {\em causal}, not {\em temporal}, structure familiar from relativistic physics. Secondly, even if these connections are ignored, we only recover temporal relations, and not spatiotemporal ones as demanded by Lewis. This, in turn, would seriously exacerbate the third objection he discusses (\citeyear[73f]{lew86}; cf.\ \S\ref{sec:unify} above), which he already took to be the by far most troubling, as we would no longer have at least spatial {\em and} temporal but only temporal relations. Therefore, the non-fundamental existence of spacetime does not save the Lewisian analysis of modality. 

\section{The relaxation to natural external relations}\label{sec:natext}

The fifth and last strategy to save Lewis's account of modality is to weaken the conditions on the worldmate relation, in the hope that the fundamental causal relations of causal set theory qualify as unifying and isolating possible worlds. The idea behind the strategy is to shed the inessential garb from the core of the Lewisian idea and to stipulate that worlds are unified and isolated by some relation satisfying a logically weaker condition. The weakening can occur in two steps: first, to ``analogically spatiotemporal'' relations and, second, to natural external relations. Let us discuss the two steps in turn. 

To recap from \S\ref{sec:unify}, {\em analogically spatiotemporal} relations are natural, pervasive, discriminating, and external in the technical meaning introduced by Lewis. These four conditions are, I take it, necessary for the analogy to succeed. Given that we assumed, for the sake of argument, that causal set theory is true of our actual world, we can safely accept that the fundamental relation is furnishes is natural. But then the troubles start. First, the condition that the relation be discriminating is only honoured thinly, in that for large causal sets, we generically expect them to fail to discriminate among all basal events: the places in the structure of relations will not be, in general, unique.\footnote{This concerns the possibility of so-called `Hegelian sets' of events. The relevant points can easily be distilled from \citep{wut12c} and \citet{wutcal17}.} 

\begin{figure}[t]
\centering
\epsfig{figure=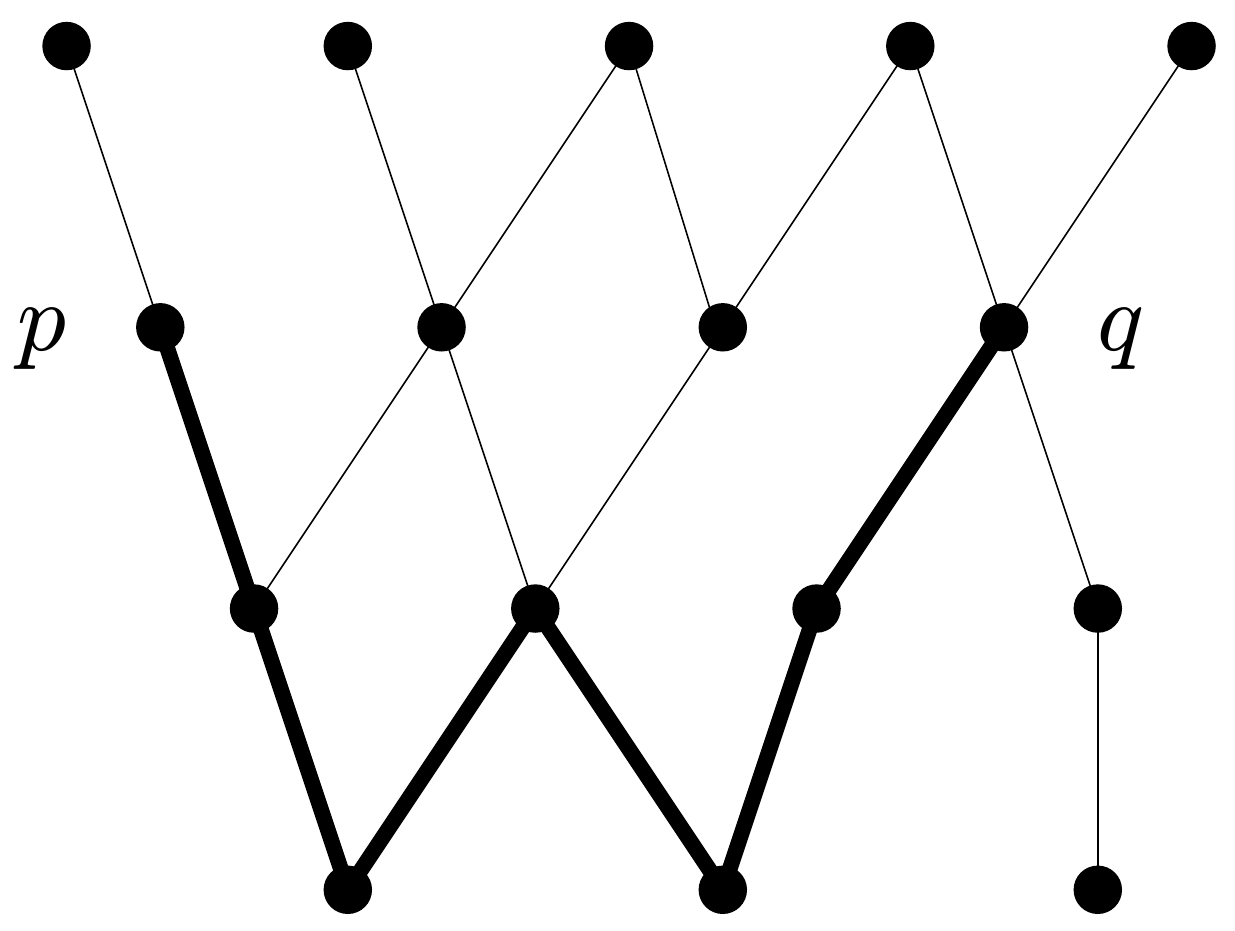,width=0.35\linewidth}
\caption{\label{fig:causalchain} As shown in this Hasse diagram of (a section of) a causal set, events $p$ and $q$ are connected by a chain of causal relations (bold lines), yet do not stand in a direct causal relation. A Hasse diagram represents the antisymmetry of the partial ordering by vertical stacking.}
\end{figure}

But the true culprit to fail the analogy is the demanded pervasiveness, which is routinely violated. Even though all events are integrated into the structure of a causal set and are typically causally linked to all other events in the causal set via long chains of direct causal relations, it is generally not the case that any given pair of them also stands in a direct causal relation, as is illustrated in Figure \ref{fig:causalchain}. 
This failure occurs for instance in `space-like related' events and results from the conjoined effects of the asymmetry of the fundamental relation and the partiality of the order it induces. These considerations suffice to see that the causal relation postulated in causal set theory do not qualify as `analogically spatiotemporal' in Lewis's.

Let us consider the second level of relaxing the conditions and stipulate that worlds are unified and isolated merely by {\em some natural external relation}, which need not be (analogically) spatiotemporal. This would maintain the Humean spirit of Lewis's proposal, as the externality of the relations permits the largely free recombination of elements of worlds.\footnote{In fact, as Cristian Mariani pointed out to me, it might be argued that the {\em only} thing that matters is that the basal entities are freely recombinable, not that the binding relation is external. That may be true, but the two aspects can no longer be cleanly separated here.} As it appeals to the naturalist, the nature of these relations may only be uncovered a posteriori. 

Here is an argument, however, to the conclusion that these relations are not external. This is surprising, as causal relations, along with spatiotemporal ones, are often taken as paradigmatic examples of external relations. The failure of externality thus warrants some unpacking. In order to do so, let us recall the basic definitions of Lewis's own classification of relations, as neatly and authoritatively summarized by Phillip Bricker:
\begin{quote}
``Things are... {\em duplicates} just in case there is a... one-to-one correspondence between their parts that preserves all natural properties and relations... An {\em intrinsic nature} is a property had by all and only the duplicates of some thing... An {\em internal} relation is a relation that supervenes on the intrinsic natures of its {\em relata}... An {\em external} relation is one that although it fails to supervene on the intrinsic natures of its {\em relata}, does supervene on the intrinsic natures of its {\em relata}, and the fusion of its {\em relata}... A relation that is either internal or external is {\em intrinsic}; all others are {\em extrinsic}.'' (\citet{bri96}, 227)
\end{quote}
Though differently motivated, Bricker also proposes a relaxation of the conditions imposed upon the worldmate relation in a Lewisian analysis of modality: ``[W]orlds are maximally {\em externally} unified regions of logical space; all and only worldmates are {\em externally} related.'' (ibid. 230) It is quite clear, then, that regardless of whether the fundamental relation in causal set theory is pervasive and discriminating, if it fails to be external, it will also wreak havoc for Bricker's suggested relaxation in Lewis's spirit. 

Let us analyze whether the fundamental relation of causal set theory is external in Lewis's own terminology. Any such analysis must consider all defensible interpretations of the metaphysics of causal set theory. I can recognize only two, leading to a dilemma for the Lewisian.

On the first horn, the basal elements have no natural properties---they are featureless after all---, in which case all of them are duplicates with the same trivial intrinsic nature. In this case, the relation supervenes neither on the relata alone, nor on the relata and their fusion together. Thus, the fundamental relation is extrinsic in Lewis's terminology on this interpretation.\footnote{One might worry, as did Valeriya Chasova in discussion with me, that if this argument succeeds, an analogous challenge could be raised against spacetime events as they feature in general relativity. While I agree that this may well imply that spatiotemporal relations as they figure in general relativity are not external on Lewis conception (though the details will sensitively depend on the metaphysics of general relativity), this will not question their {\em spatiotemporality}.} This may be considered too quick: why should the Lewisian accept that the mereological sum of a pair of events standing in this relation could have a duplicate such that the parts of that sum do {\em not} stand in the relation?\footnote{I thank George Darby for pushing this point. Lewis's ``composite'' (1986, 62) and Bricker's ``fusion'' (see the quote above) are rendered as `mereological sum' here.} This would amount to setting the sum or fusion of the relata identical to the subset of basal elements at stake, together with the full causal substructure induced by the entire causal set on the subset. However reasonable this interpretation may be, it seems to deny its own presupposition on this horn of the dilemma, viz., that the basal elements have no natural properties. 

Alternatively for the second horn, the basal elements' relational profile constitutes their natural properties---this is what identifies them after all---, in which case only elements in the same position in isomorphic structures will be duplicates. In this case, the relation supervenes on the relata alone and is thus internal in Lewis's own terminology. 

Either way, the relation is not external, thwarting Lewisian and Brickerian attempts to relax the requirements for the worldmate relation in the face of causal set theory. The dilemma shows just how precarious any via media between the Scylla of making the relations extrinsic and the Charybdis of rendering them internal must be in the case of a metaphysically simple---almost impoverished---fundamental structure. So any successful relaxation will have to go beyond Bricker's already rather weak conditions. In fact, the only remaining stand for the Lewisian before the abyss of triviality seems to just require the worldmate relation to be defined in terms of the natural fundamental relations procured by our most fundamental theory. For the causal sets investigated here, this would mean to just use the fundamental causal relation delivered by the theory. While perfectly naturalistic in spirit, this move would imply a decisive rapprochement between metaphysical and physical modality. In fact, the difference between the two would arguably degenerate into something like the difference between kinematic and dynamical possibilities in the physical theory. Desirable or altogether unpalatable, I leave these implications for another occasion.

Also to a future occasion will be deferred the consequences of the disappearance of spacetime for Lewis's ontology, and for Humean supervenience and laws of nature more generally. As Lewis fundamentally admits basically only spacetime events as particulars, the perfectly natural monadic properties they instantiate, and the perfectly natural relations which they exemplify \citep[\S2]{hal10}, his fundamental ontology stand in obvious tension with a theory denying fundamental status to spacetime. And while many Humeans may not care too much about Lewis's fundamental ontology---let alone his modal realism---, they will be challenged to recast Lewis's formulation of Humean supervenience in non-spatiotemporal terms in their attempts to articulate metaphysical theories of laws of nature and of modality consonant with the Humean spirit.

\section{Summary and outlook}\label{sec:conc}

In conclusion, if a true fundamental theory such as causal set theory ruled out spacetime as being ontological part and parcel of the fundamental reality described by the theory, and if it was shown how spacetime emerges as an approximation from the fundamental structure thus connecting the theory to known physics and rendering it empirically coherent, then our actual world is not possible according to Lewis's analysis of modality. Or at least if this analysis is supposed to regiment more than just emergent modality. Of course, the Lewisian could content herself with a theory of modality restricted to the emergent level, thus covering almost all of human modal discourse. Nothing of what I said precludes this restriction, but suffice it to say that such a restricted analysis would neither be globally applicable as to include our theorizing about fundamental reality nor be based on a fundamental and true understanding of our world. For both these reasons, the move to emergent modality reeks of desperation. I see no principled reason why we couldn't do better than that.

The conclusion that on Lewis's analysis, the actual world is impossible only depended on the assumption that according to the most fundamental theory of gravity spacetime does not form part of the furniture of the world and on what I take to be a mild form of naturalism, viz.\ that the fundamental structure of our world is best described by our best physical theories. Obviously, my conclusion can be resisted by repudiating either of these premises. However, the first assumption is supported---but of course not guaranteed---by recent developments in fundamental physics, and in particular in quantum gravity. Given that there exists the serious possibility that spacetime is absent from fundamental reality, I hope you agree, esteemed reader, that it is worth considering its metaphysical consequences, at least if you share my mild naturalism. I hope to have shown that these consequences may be significant and far-reaching.

\bibliographystyle{plainnat}
\bibliography{/Users/christian/Professional/Bibliographies/quantumgravity}

\begin{thebibliography}{19}
\providecommand{\natexlab}[1]{#1}
\providecommand{\url}[1]{\texttt{#1}}
\expandafter\ifx\csname urlstyle\endcsname\relax
  \providecommand{\doi}[1]{doi: #1}\else
  \providecommand{\doi}{doi: \begingroup \urlstyle{rm}\Url}\fi

\bibitem[Barrett(1999)]{bar99}
Jeffrey~A Barrett.
\newblock \emph{The Quantum Mechanics of Minds and Worlds}.
\newblock Oxford University Press, 1999.

\bibitem[Bell(1987)]{bel87}
John~S Bell.
\newblock \emph{Speakable and Unspeakable}.
\newblock Cambridge University Press, 1987.

\bibitem[Bricker(1996)]{bri96}
Phillip Bricker.
\newblock Isolation and unification: The realist analysis of possible worlds.
\newblock \emph{Philosophical Studies}, 84:\penalty0 225--238, 1996.

\bibitem[Hall(2016)]{hal10}
Ned Hall.
\newblock David {Lewis's} metaphysics.
\newblock In Edward~N. Zalta, editor, \emph{Stanford Encyclopedia of
  Philosophy}, 2016.
\newblock URL \url{https://plato.stanford.edu/entries/lewis-metaphysics/}.

\bibitem[Hawking et~al.(1976)Hawking, King, and McCarthy]{haweal76}
Stephen~W Hawking, A~R King, and P~J McCarthy.
\newblock A new topology for curved space-time which incorporates the causal,
  differential, and conformal structures.
\newblock \emph{Journal of Mathematical Physics}, 17:\penalty0 174--181, 1976.

\bibitem[Huggett and W\"uthrich(2013)]{hugwut13b}
Nick Huggett and Christian W\"uthrich.
\newblock Emergent spacetime and empirical (in)coherence.
\newblock \emph{Studies in the History and Philosophy of Modern Physics},
  44:\penalty0 276--285, 2013.

\bibitem[Huggett and W\"uthrich(forthcoming)]{hugwut}
Nick Huggett and Christian W\"uthrich.
\newblock \emph{Out of Nowhere: The Emergence of Spacetime in Quantum Theories
  of Gravity}.
\newblock Oxford University Press, Oxford, forthcoming.

\bibitem[Huggett et~al.(2013)Huggett, Vistarini, and W\"uthrich]{hugeal13}
Nick Huggett, Tiziana Vistarini, and Christian W\"uthrich.
\newblock Time in quantum gravity.
\newblock In Adrian Bardon and Heather Dyke, editors, \emph{A Companion to the
  Philosophy of Time}, pages 242--261. Wiley-Blackwell, Chichester, 2013.

\bibitem[Lam and W\"uthrich(forthcoming)]{lamwut18}
Vincent Lam and Christian W\"uthrich.
\newblock Spacetime is as spacetime does.
\newblock \emph{Studies in the History and Philosophy of Modern Physics},
  forthcoming.

\bibitem[Lewis(1986)]{lew86}
David Lewis.
\newblock \emph{On the Plurality of Worlds}.
\newblock Blackwell, 1986.

\bibitem[Malament(1977)]{mal77}
David~B Malament.
\newblock The class of continuous timelike curves determines the topology of
  spacetime.
\newblock \emph{Journal of Mathematical Physics}, 18:\penalty0 1399--1404,
  1977.

\bibitem[Maudlin(2007)]{mau07}
Tim Maudlin.
\newblock Completeness, supervenience, and ontology.
\newblock \emph{Journal of Physics A: Mathematical and Theoretical},
  40:\penalty0 3151--3171, 2007.

\bibitem[Rideout and Wallden(2009)]{ridwal09}
David Rideout and Petros Wallden.
\newblock Spacelike distance from discrete causal order.
\newblock \emph{Classical and Quantum Gravity}, 26:\penalty0 155013, 2009.

\bibitem[Schaffer(2009)]{sch03}
Jonathan Schaffer.
\newblock Is there a fundamental level?
\newblock \emph{No\^{u}s}, 37:\penalty0 498--517, 2009.

\bibitem[Smolin(2009)]{smo09}
Lee Smolin.
\newblock Generic predictions of quantum theories of gravity.
\newblock In Daniele Oriti, editor, \emph{Approaches to Quantum Gravity: Toward
  a New Understanding of Space, Time and Matter}, pages 548--570. Cambridge
  University Press, Cambridge, 2009.

\bibitem[Witten(1996)]{wit96}
Edward Witten.
\newblock Reflections on the fate of spacetime.
\newblock \emph{Physics Today}, pages 24--30, April 1996.

\bibitem[W\"uthrich(2012)]{wut12c}
Christian W\"uthrich.
\newblock The structure of causal sets.
\newblock \emph{Journal for General Philosophy of Science}, 43:\penalty0
  223--241, 2012.

\bibitem[W\"uthrich and Callender(2017)]{wutcal17}
Christian W\"uthrich and Craig Callender.
\newblock What becomes of a causal set?
\newblock \emph{British Journal for the Philosophy of Science}, 68:\penalty0
  907--925, 2017.

\bibitem[Zorn(1935)]{zor35}
Max Zorn.
\newblock A remark on method in transfinite algebra.
\newblock \emph{Bulletin of the American Mathematical Society}, 41:\penalty0
  667--670, 1935.

\end{thebibliography}

\end{document}